\documentclass[aps,twocolumn,prl,showpacs]{revtex4}
\usepackage{natbib}

\newcommand{\grl}{GeoRL}

\newcommand{\jgr}{J.~Geophys.~Res.}
\usepackage{graphicx}
\usepackage{graphics}
\begin{document}
\bibliographystyle{apsrev}

\title[Solar Forcing of the Streamflow]{Solar Forcing of the Streamflow of a Continental Scale South American River}



\author{Pablo J. D. Mauas}
\email{pablo@iafe.uba.ar}
\affiliation{Instituto de Astronom\'\i a y F\'\i sica del Espacio
  (CONICET-UBA),  C.C. 67 Sucursal 28, 1428,Buenos Aires, Argentina}

\author{Eduardo Flamenco}
\affiliation{Instituto Nacional de Tecnolog\'\i a Agropecuaria,
  Rivadavia 1439, 1033,  Buenos
  Aires, Argentina}

\author{Andrea P. Buccino}
\affiliation{Instituto de Astronom\'\i a y F\'\i sica del Espacio,(CONICET-UBA),  C.C. 67 Sucursal 28, 1428,Buenos Aires, Argentina}


\date{\today}

\begin{abstract}

Solar forcing on climate has been reported in several studies although
 the evidence so far  remains inconclusive. Here, we analyze the
 streamflow of one of the largest rivers in the world, the Paran\'a in
 southeastern South America. For the last century, we find a strong
 correlation with Sunspot Number, in multi-decadal time scales, with larger
 solar activity corresponding with larger streamflow. The correlation
 coefficient is r=0.78, significant to a 99\% level. In shorter timescales we
 find a strong correlation with El Ni\~no. These results are a step
 toward flood prediction, which might have large social and economic impacts. 

\end{abstract}

\maketitle


\section{Introduction}
Evidence of a solar influence on climate has been traditionally found 
on records of Northern Hemisphere temperature
\citep{1991Sci...254..698F}, or sea surface temperature
\citep{1997JGR...102.3255W}, usually related to
changes in solar irradiance \citep{2006JGRC..11109020W, 2008JGRC..11301002W}. 
Also, a link between solar activity and 
cloud cover through the action of cosmic rays has been proposed
\citep{2000PhRvL..85.5004M} and widely discussed
\citep{2002JGRD..107.4211S}. Recently, a number of studies took a new
approach to the problem, looking into paleoclimatic records of
atmospheric moisture. For example, evidences were found of the solar
influence on the Asian Monsoon \cite{2001Natur.411..290N,
2002E&PSL.198..521A, 2003Natur.421..354G, 2003Sci...300.1737F,
2005Sci...308..854W}, 
in the drought conditions in Africa \citep{2000Natur.403..410V} and
Mexico \citep{2001E&PSL.192..109H}, and in general in tropical
precipitation regimes \citep{2004JASTP..66.1767V}. The influence of
solar activity on regional precipitations was also found in
experiments with a global climate model \citep{2003JCli...16..426M}.


Here, we take a different approach to the problem, looking into a
different climatic variable, also related to moisture, in a different
time scale: we study the streamflow of the Paran\'a River during the
last 100 years \citep{2005MmSAI..76.1002M}.

River streamflows are excellent climatic indicators since they
integrate precipitations, infiltrations and evapotranspiration over
large areas.  In particular, those rivers with continental scale
basins smooth out local variations, and can be particularly useful to
study global forcing mechanisms. Moreover, knowledge and/or prediction
of streamflow regimes is fundamental for different social and economic
reasons, from the prediction of floods and droughts to planning of
agricultural or hydroenergetic conditions. 

\section{Data}
The Paran\'a is one of the largest rivers in the world: with a basin
area of over 3,100,000 km$^2$ and a mean streamflow, during the last two
decades of the 20$^\textrm{th}$ century, of 20,600 m$^3$/s, the Paran\'a is the fifth
river of the world according to drainage area and the fourth according
to streamflow. With its origin in the southernmost part of the Amazon
forest, it flows south collecting water from the countries of Brazil,
Paraguay, Bolivia, Uruguay and Argentina, and forms one of the
mightiest deltas of the world before its outlet in the Plata River, a
few kilometers north of the City of Buenos Aires. Due to the fact
that, unlike other rivers of similar size like the Amazon or the
Congo, it flows through heavily populated areas, and that it is
navigated by overseas trade ships, it has one of the longest
streamflow data series, which covers the last century.

\begin{figure}[htb!]
\centering
\includegraphics[width=0.45\textwidth]{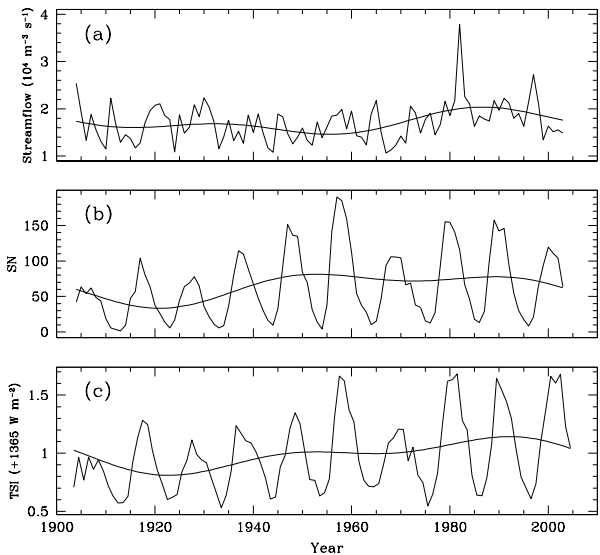}
\caption{(a) Paran\'a's annual streamflow at the Corrientes gauging
 station. (b) Yearly International Sunspot Number ($S_N$). (c) Solar
 irradiance reconstruction \cite{2005ApJ...625..522W}.  
 The secular trends, obtained with a low-pass Fourier filter with a 50
 years cut-off, are shown as thick lines.}\label{fig_1} 
 \end{figure}

Here we analyze the streamflow data measured at a gauging station
 located in the city of Corrientes, 900 km north of the outlet of the
 Paran\'a. It is measured continuously from 1904, on a daily
 basis. The Paran\'a's hydrological year goes from September to August,
 with maximum streamflow in the (Southern Hemisphere's) summer months
 of January, February and March. We therefore build our yearly series
 integrating the flow from September to August of the next year. The
 data are shown in Fig. \ref{fig_1}(a), together with the trend obtained with a
 low-pass Fourier filter with a 50 years cut-off.

It can be seen that the flow of the Paran\'a is larger in the last three
 decades, with a mean value almost 20\% larger than that of the first
 seventy years of the 20$^\textrm{th}$ century. In particular, the streamflow
 during the last 30 years has increased in the months in which the
 flow is minimum, May to December, while the flow remains more or less
 constant during the months of maximum. This trend has already been
 noticed, and was attributed to Amazonian deforestation \citep{1991JCli....4..957N}, which
 should facilitate water drainage. However, the same trend is also
 found in other rivers of the region like the Iguaz\'u, whose sub-basin
 has not undergone significant changes in land use during the
 20$^\textrm{th}$ Century \citep{Garcia98}. On the other hand, this trend can be
 considered as an integral part of the large-scale variations of the
 climate system \citep{1998JCli...11.2858G}. It should be noted that southeastern South
 America is one of the principal regions of the globe where land
 surface temperature has been increasing since 1900 \citep{1998JCli...11.2570R}.

As a solar activity indicator we consider the yearly Sunspot Number
($S_N$) \cite{sunspot}, which is shown in Fig. \ref{fig_1}(b) together with
its trend, obtained in the same way than that for the streamflow. 
Alternatively, the irradiance reconstruction by Wang et al.
\cite{2005ApJ...625..522W} can be used as a solar activity
indicator. It differs somehow from the sunspot record since irradiance
depends more on bright regions on the solar surface. 

\begin{figure}[htb!]
\resizebox{\hsize}{!}{ \includegraphics{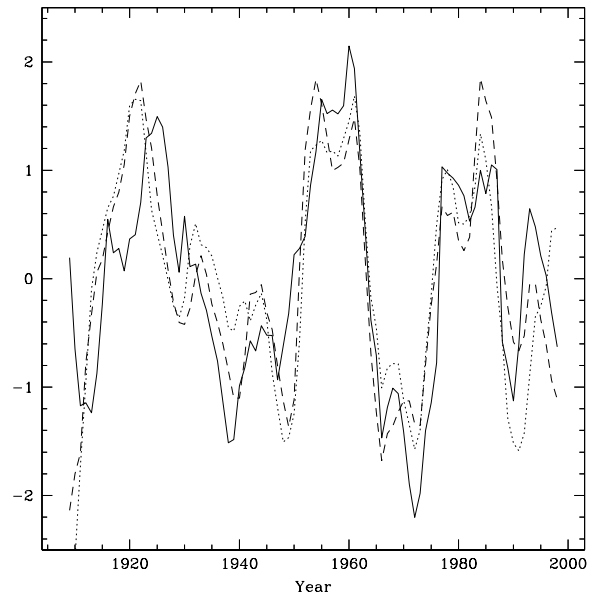}}
 \caption{The detrended time series for the Paran\'a's streamflow
 (full line), the sunspot number (dashed line) and the irradiance
 reconstruction (dotted line). The detrended
 series were obtained by subtracting from each data series the
 corresponding trend, shown as a thick line in Fig. \ref{fig_1}, and were
 smoothed by an 11-yr running mean to smooth out the solar cycle. The
 series were normalized by subtracting the mean and dividing by the
 standard deviation, to avoid introducing arbitrary free parameters. 
 The Pearson's correlation coefficient of the streamflow data with the
 sunspot number is r=0.78, and with the irradiance reconstruction is
 r=0.69.} \label{fig_2} 
\end{figure}

In \mbox{Fig. \ref{fig_2}} we show the detrended time series for streamflow,
$S_N$ and the irradiance reconstruction. In all cases we have
substracted the trend shown in 
Fig. \ref{fig_1} from the annual data, and we have performed an 11
year running-mean to smooth out the solar cycle. When plotting
together different quantities, two free parameters are usually
introduced, namely, the offset and the relative scales. To avoid these
two artificial parameters, we have normalized the three quantities by
substracting the mean and dividing by the standard deviation of each
series.

\section{Results}

Visual agreement between the Paran\'a's streamflow, the sunspot number
and the irradiance reconstruction shown in \mbox{Fig. \ref{fig_2}} is
quite remarkable. The Pearson's correlation coefficient between
streamflow and SN is r=0.78, and between streamflow and irradiance is
r=0.69. We performed a t-Student test to check the significance of
these correlations, reducing the 
number of effective points to take into account the autocorrelation of
the series and the smoothing, and we obtained a significance level
higher than 99.99\% in both cases.  We also analyzed the significance
level of the
correlations with a non-parametric random-phase test
\cite{1997JCli...10.2147E}. For the SN series, we obtained that only a
0.4\% of the 
random series presented a correlation coefficient greater than
0.78. In this way, we obtain a significance level of 99.6\% for the
correlation between streamflow and SN.
For the irradiance reconstruction, the significance obtained with this
method is larger than 99.99\%.

Recently, a mechanism has been proposed for the influence of solar
activity on climate, involving the modulation of galactic cosmic rays
(GCR) by the interplanetary magnetic field associated with the solar
wind and, therefore, with solar activity. In this picture, GCR would
affect  cloud formation on earth, through ionization of the
terrestrial atmosphere. Therefore, periods of higher solar activity,
when the interplanetary magnetic field is larger, and therefore less
GCR hit the earth, the cloud cover would be smaller. For this reason,
it is particularly interesting to check whether there is a
particularly strong correlation between the Paran\'a's discharge and
GCR. 


Therefore, we have also checked the correlation with two other solar-activity
indexes. First, we considered the neutron count at Climax, Colorado,
available since 1953 \citep{climax}. Since neutrons are produced
when GCR hit the upper atmosphere, neutron count is a direct measure of GCR
flux. Furthermore, since GCR flux in different parts of the world
depend only on latitude, following the strength of the terrestrial
magnetic field, Climax's values are representative of GCR flux everywhere.  

The other index we used was the aa index, which is a
measure of the disturbance level of the Earth's magnetic field based
on magnetometer observations of two, nearly antipodal, stations in
Australia and England \citep{1998GeoRL..25.1035C}, and it is available
since 1868 \cite{geomag}. It is 
worth pointing out that the aa index follows the envelope of solar
activity, and while $S_N$ returns to zero at each solar minimum, aa
minima reflect the long-term level of solar activity seen in
Fig. \ref{fig_1}(b). Since the Earth's magnetic field, which is affected by the
solar wind, determines how much of the GCR flux ultimately reaches the
Earth, aa can also be used to test the GCR-climate hypothesis.  

In both cases, we found a correlation with Paran\'a's streamflow, as
expected since all indexes of solar activity are correlated between
each other. However, the correlations shown in Fig. \ref{fig_2} are the
largest, pointing to a more direct correlation with solar irradiance
than with GCR.  


An important point to be stressed regards the sign of the relationship
between solar activity and river discharge reported here, which
implies that wetter conditions in this area coincide with periods of
higher solar activity. This is in agreement with paleoclimatic studies
of the Asian monsoon \cite{2002E&PSL.198..521A,2003Natur.421..354G,vonrad} which report an increase in monsoon
during periods of increased solar activity. Also, increases in solar
activity were found to be correlated with increased moisture over
Alaska during the Holocene \citep{2003Sci...301.1890H}, and similar results were found in
simulations of climate during a period of reduced solar activity known
as the Maunder Minimum \citep{2001Sci...294.2149S}.

In contrast, studies in East Africa
report severe droughts during phases of high solar activity and
increased precipitation during periods of low solar
irradiation \citep{2000Natur.403..410V}.  To explain these differences
it has been proposed that increased 
solar irradiation causes more evaporation in equatorial regions,
enhancing the net transport of moisture flux to the Indian
sub-continent via monsoon winds \citep{2002E&PSL.198..521A}. A similar
mechanism was found in 
simulations with a climate model, in which enhanced solar forcing
produces greater evaporation in relative cloud-free regions in the
subtropics, and the resulting moisture then converges into the
precipitation convergence \mbox{zones \citep{2003JCli...16..426M}.}  

In the American continent,
droughts in the Yucatan Peninsula have been associated with periods of
high solar activity and even proposed to explain the Mayan
decline \citep{2001E&PSL.192..109H}, in contrast with the results found
here. This is in agreement 
with an inverse correlation that was found between the southern and
northern regions of South America, with dry periods in the South
corresponding almost in phase to humid intervals in the North and \mbox{vice
versa \citep{Iri99}.}  

The fact that solar influence is different in different
parts of the world is of particular importance when assessing the
proposed relationship between solar activity and climatic change,
since it points out to modifications in circulation patterns or other
mechanisms that do not globally affect climatic variables like
moisture, but affect their distribution instead.

\begin{figure}[htb!]
\centering
\includegraphics[width=0.395\textwidth]{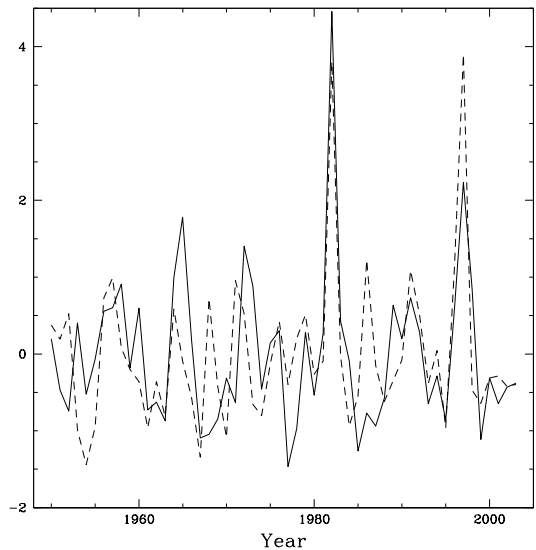} 
\caption{Paran\'a's streamflow minus the multi-decadal component
 (full line) compared with the Ni\~no1+2 index (dotted line), which is
 available from 1950. Both series were normalized by subtracting the
 mean and dividing by the standard deviation, to avoid introducing
 arbitrary free parameters. The Pearson's correlation coefficient for
 the whole series is r=0.65.}\label{fig_3} \end{figure}

Finally, in Fig \ref{fig_3} we show the high frequency variations of
the Paran\'a's streamflow, obtained by subtracting the
11-year-running-mean from the yearly data. Also shown is the Ni\~no1+2
index, averaged from September to August next year, to coincide with
the hydrological year of the river. El Ni\~no1+2, which is available
since 1950, is a measure of the sea surface temperatures in the
Equatorial Pacific Ocean, close to the South American coast
(0-10$^\textrm{o}$ S, 80-90$^\textrm{o}$ W) \cite{indices}.  In this
case, a very good accordance between both curves can be seen, with a
correlation coefficient r=0.65 and a significance larger than
99.99\%. In particular, the large annual discharges of 1982 and 1997
are associated with two exceptional El Ni\~no episodes.

This relation between the Paran\'a's streamflow and the ENSO
phenomenon and, in particular, the sea surface temperatures in a
region of the tropical Pacific, was already used in flood predictions
in this basin \citep{Flamenco}. A similar correlation was also found
for the Indian monsoon~\citep{1997GeoRL..24..159M}.

\section{Conclusions}

Streamflow variability of the Paran\'a river has three
temporal components: on the secular scale, it is probably part
of the global climatic change, which at least in this region of the
world is related with more humid conditions; on the  multi-decadal time
scale, we found a strong correlation with solar activity, as expressed
by the Sunspot Number, and therefore probably with solar irradiance,
with higher activity coincident with larger discharges; on the yearly
time-scale, the dominant correlation is with El Ni\~no.  

These correlations can be used for flood prediction: a regression
between Paran\'a's streamflow ($S$), the sunspot number filtered as in
Fig. \ref{fig_2} ($S_{Nf}$), and El Ni\~no1+2 ($N_{12}$), gives:
\begin{displaymath}
S=(63.6\pm59.3)\,S_{Nf}+(3.6\pm0.6)\times10^{2}\,N_{12}+(6.5\pm1.3)\times10^4
\end{displaymath}
between 1949 and 1999, with a Pearson correlation coefficient
r=0.66, significant to 99.99\%.

Early flood prediction, in fact, has large social
and economic impacts:~During the last flood, in 1997, 180,000~km$^2$ of
land were covered with water, 125,000 people have to be evacuated, and
25 people died. In all, the three largest floods of the Paran\'a during
the 20$^\textrm{th}$ century caused economic losses for five thousand million
dollars.





%
%

%
%

\end{document}